# Lost Waterways: Clues from Digitized Historical Maps of Manila and other Philippines Cities

Karol Giuseppe A. Jubilo[1], Meryl Regine L. Algodon[1], Elexis Mae A. Torres[1], Zayda Domini P. Abraham, Ari Ide-Ektessabi[2], Maricor N. Soriano[1*]

[1]National Institute of Physics, University of the Philippines Diliman

Quezon City 1101, Philippines

[2]Advanced Imaging Technology, Graduate School of Engineering, Kyoto University

*`msoriano@nip.upd.edu.ph`

**Abstract.** We search for lost bodies of water in the cities of Manila, Tacloban, Iloilo, Cebu, Davao, and Naga by aligning their digitized Spanish-era and American-era maps to Google maps. These vanished ancient waterways can either become flooding hazards in case of extreme weather events, or liquefaction hazards, in case of earthquakes. Digitized historical maps of the cities were georectified, overlaid on current Google maps, and checked for potential missing bodies of water. Inspection through field visits and interviews with locals were conducted to verify the actual status of suspected sites. The validation identified lost, found, and even new bodies of water. There was also evidence of affected buildings, rainless flooding, and a "new normal" for the meaning of flooding among frequently inundated residents.

**Keywords.** Historical maps, flooding hazards, bodies of water image processing algorithms, image transformations,

## 1 Introduction

Inland bodies of water are extremely important resources that, even if cartographers from centuries ago do not have access to aerial imagery, most likely they will depict the location of lakes and rivers with fair accuracy. Through time some of these bodies of water would have vanished due to natural causes (e.g. sedimentation) or human activity (e.g. reclamation). Erecting structures on top of these lost bodies of water may be hazardous as these areas are prone to flooding after severe rainfall [01], or to liquefaction and amplified ground shaking during an earthquake.

In this work we aligned digitized 19th century and 1940's maps of six Philippine cities with Google Map to determine areas where original bodies of water have vanished. We then use Google Street View to inspect if the bodies of water have indeed disappeared or were simply left out in the Google map. We visited and inspected each of the sites and interviewed locals about flooding incidents. By identifying lost waterways we can create hazard maps that can be used for disaster mitigation especially when Metro Manila and the rest of the Philippines are highly disaster-prone [03, 04].

Overlaying historical maps with present maps has been done by other groups for different purposes such as historic visualization [05], urban change analysis, [07,09], and understanding environmental issues [06, 08]. Our work extends the purpose to disaster mitigation by providing a map of potential flooding/liquefaction hazards down to the street level. Like previous work we can only reliably georeference historical maps by using streets, roads, and other built structures that have remained up to the present as ground control points [02].

## 2   Methodology

### 2.1 Digitization

Historical maps were sourced from the National Archives of the Philippines, museums, libraries, and online repositories. Hard copies of maps were digitized using the NIJI-S Scanner developed by the Advanced Imaging Science and Technology Kyoto University. The NIJI-S uses cold light (white LEDs) for illumination and a line scan camera for capture and is custom-built for digitizing fragile 2D cultural heritage objects such as old paintings and historical documents. Figure 1 shows our team performing map digitization in the School of Urban and Regional Planning Library using NIJI-S.

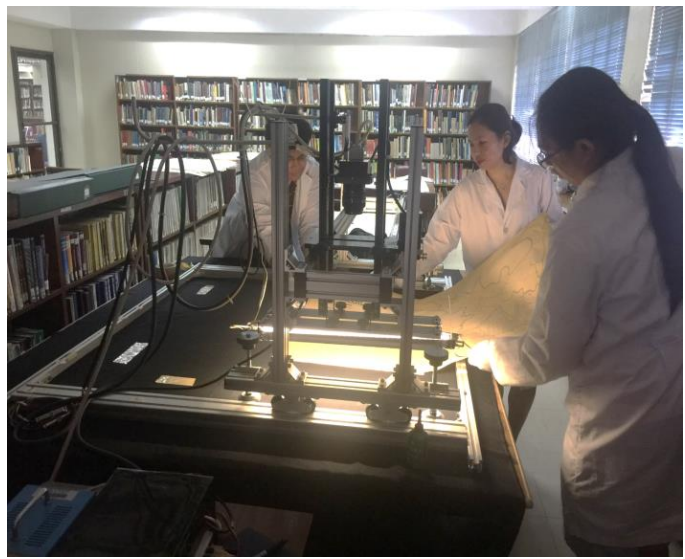

*Figure 1. Map digitization using the NIJI-S scanner.*

### 2.2 Georectification

We used MATLAB Mapping Toolbox to rectify the historical maps of the cities in the Philippines onto their corresponding Google maps. The toolbox uses the function `cpselect` to allow manual selection of common points (known as Ground Control Points or GCPs) on both the historical map and the Google map. The function `fitgeotrans` creates a transformation matrix from the set of selected GCPs. The function requires an input of a transformation type. We applied projective and second order polynomial transformation types in this study. The warping and rectification of the historical map is performed by the `imwarp` function using the generated transformation matrices. The rectified historical map was then overlaid on Google map.

The resulting overlaid map was used to compare the historical maps with the Google maps to look for missing bodies of water. The missing bodies of water were confirmed via Google Street View and fieldwork validation.

## 3   Scope of Work

Although scanning, georeferencing, and change analysis was done on several maps, we limited our fieldwork validation to six major cities in the Philippines namely: Manila City, Iloilo City, Tacloban City, Davao City, Cebu City, and Naga City. Table 1 lists the fieldwork dates per city. The objective of the fieldwork is to validate the locations of the bodies of water that disappeared according to the processed map. The teams took photographs of the specified locations and interviewed local officials and residents about flooding incidents.

**Table 1.** List of cities visited and their corresponding fieldwork schedules.

| City | Date |
|---|---|
| Manila City | August 2, 2017 |
| Iloilo City | November 12-16, 2017 |
| Tacloban City | November 13-17, 2017 |
| Davao City | November 12-17, 2017 |
| Cebu City | March 8 and June 10-13, 2017 |
| Naga City | December 10-13, 2017 |

## 4   Results and Discussion

### 4.1 Manila

Figure 2 shows two historical Manila maps (1899 and 1908) along with a recent Google map where the Pasig river and its estuaries cut through. Through time several small estuaries have disappeared. The lower leftmost body of water in the 1899 map disappeared in the 1908 and Google map. Another estuary present in the middle upper portion of the 1899 and 1908 maps disappeared completely in the Google maps.

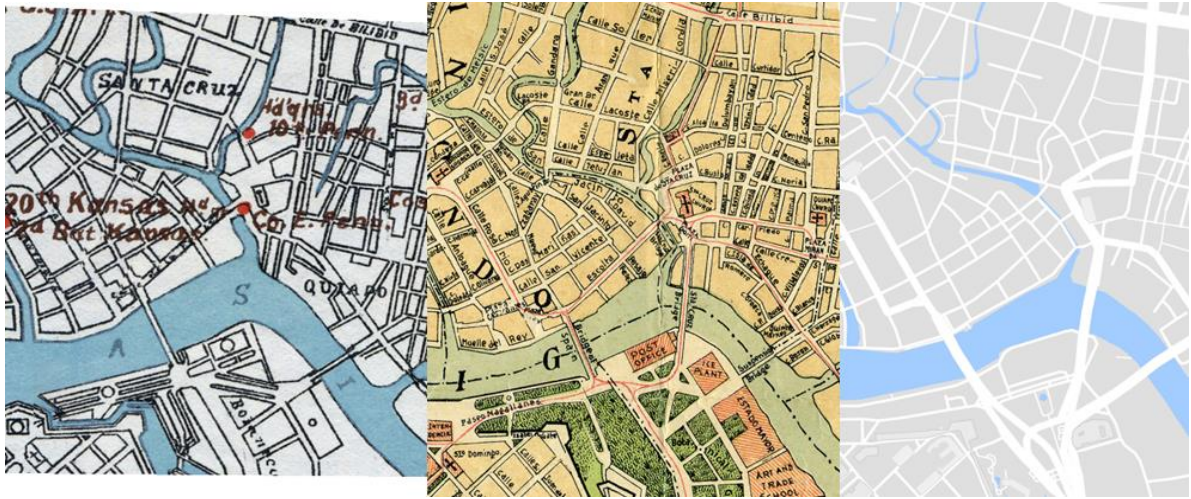

*Figure 2. Georeferenced maps of Manila showing a portion of Pasig River and its estuaries. Left - 1899 map courtesy of the Filipinas Heritage Library, Center- 1908 map courtesy of the Filipinas Heritage Library, Right - Google Map.*

The validation fieldwork team discovered that an exposed culvert (Figure 3) can be seen on top of estuary that disappeared in the 1908 and Google maps. This opens the possibility that estuaries that are absent in Google map could have been channeled underground.

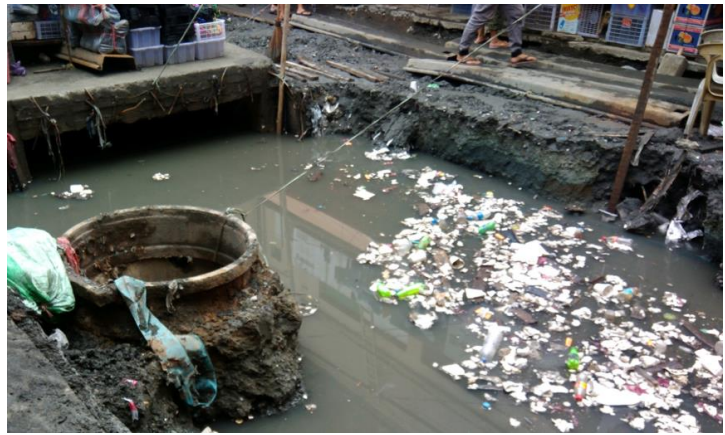

*Figure 3 An exposed culvert discovered by the validation fieldwork team*

In many cases during the validation fieldwork, the interviewed locals claim they do not experience flooding in their area. Yet they describe that the water level does rise during rain but lowers just as fast. If the water does not linger after the rain, and is just ankle-deep, they do not consider it flooding. It may be possible that the locals have already developed a different baseline for the meaning of flooding.

Traced in Figure 4 is a map summary of field surveyed Pasig River estuaries that are confirmed to have vanished (red), are still existing but not present in Google map (green) or may have been diverted underground (violet).

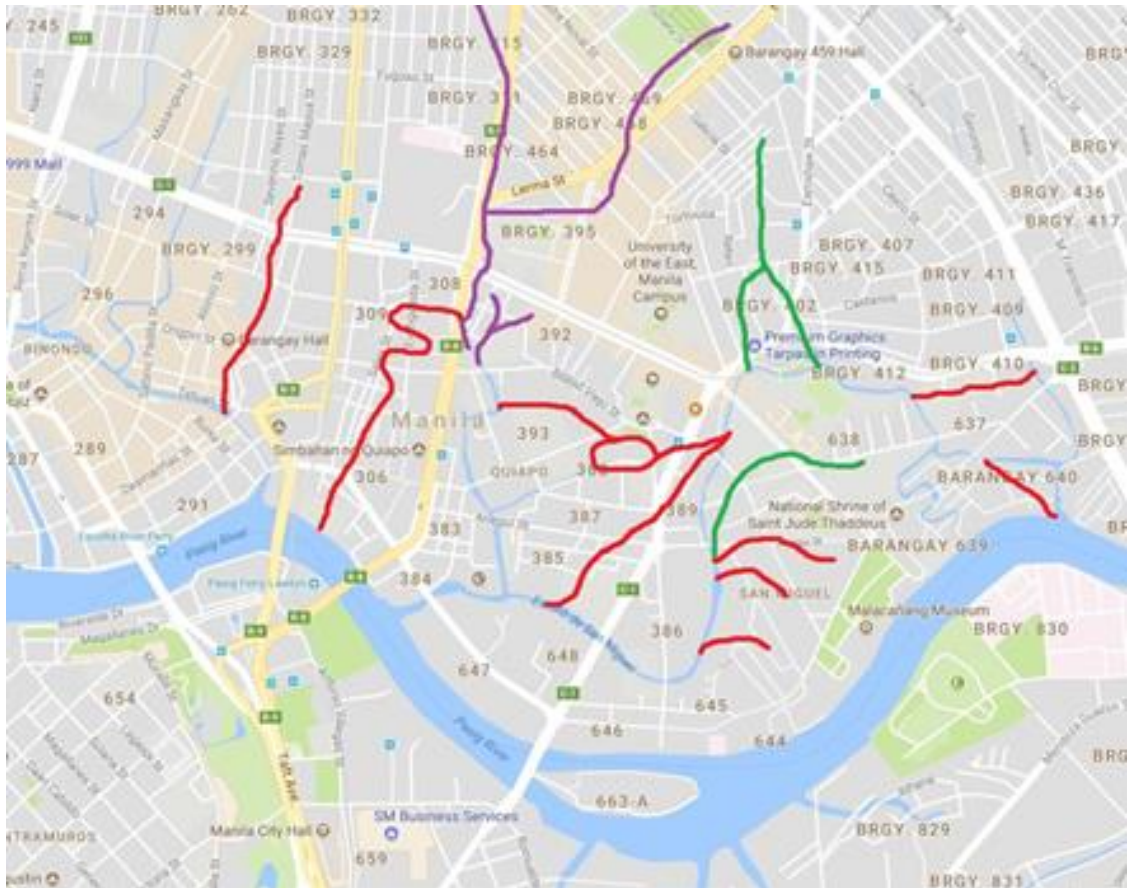

*Figure 4. Detected changes in waterways around Manila, Red-disappeared, Green- still present, Violet– possibly underground.*

## 4.2 Iloilo

Figure 5 shows a portion of the 1833 Iloilo map and its corresponding Google map. The team observed that the local government was installing big box culverts (shown in Figure 6) on areas where old tributaries of the Iloilo river were located. According to local officials the area is a high flooding risk and the civil works were part of their continuing effort to improve the drainage system. In contrast to flooding risks, community wells like the one pictured in Figure 7 were found in the area where the waterways disappeared, an indication of a freshwater source underground.

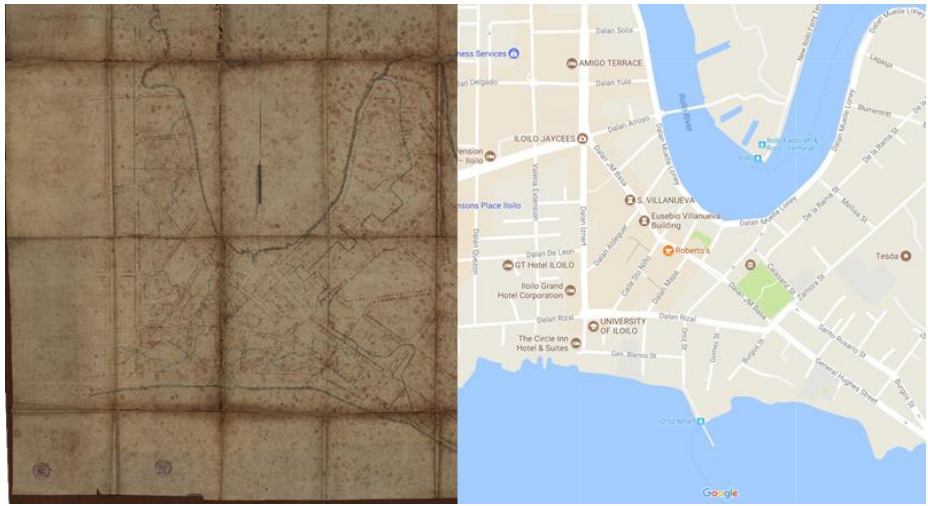

*Figure 5 Left - Portion of 1833 Iloilo City map courtesy of the National Archives of the Philippines, Right - corresponding Google Map.*

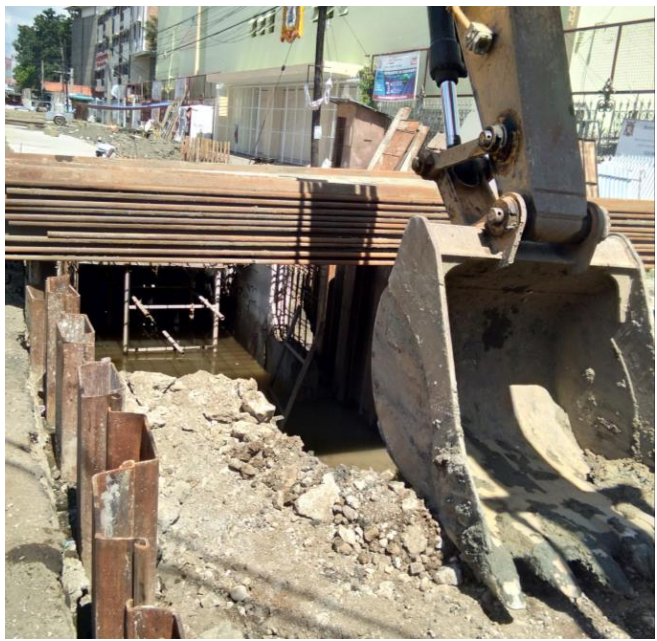

*Figure 6. Culverts in Iloilo City being installed in a high flood risk area.*

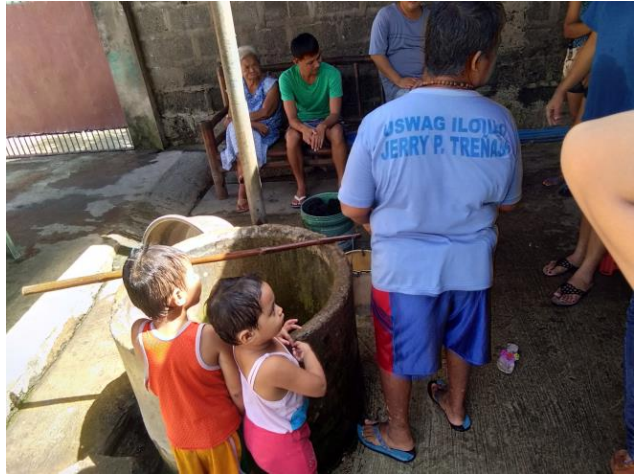

*Figure 7. A community freshwater well was found in the vicinity of an old natural pond.*

During a courtesy call with the Iloilo City mayor, the fieldwork team was also informed about a land lot along Dalan Montinola that was identified by our team to be of high risk to liquefaction. According to the mayor's personal account, a building, more than five stories high, was built on the said lot, however, only after a few months, the building was observed to be tilting. Construction was then halted until thorough inspection was made on the property. Based on the 1833 map, the present Dalan Montinola appears to be a backwater of the Iloilo River.

Figure 8 shows the visited sites in Iloilo City that were confirmed to have vanished bodies of water.

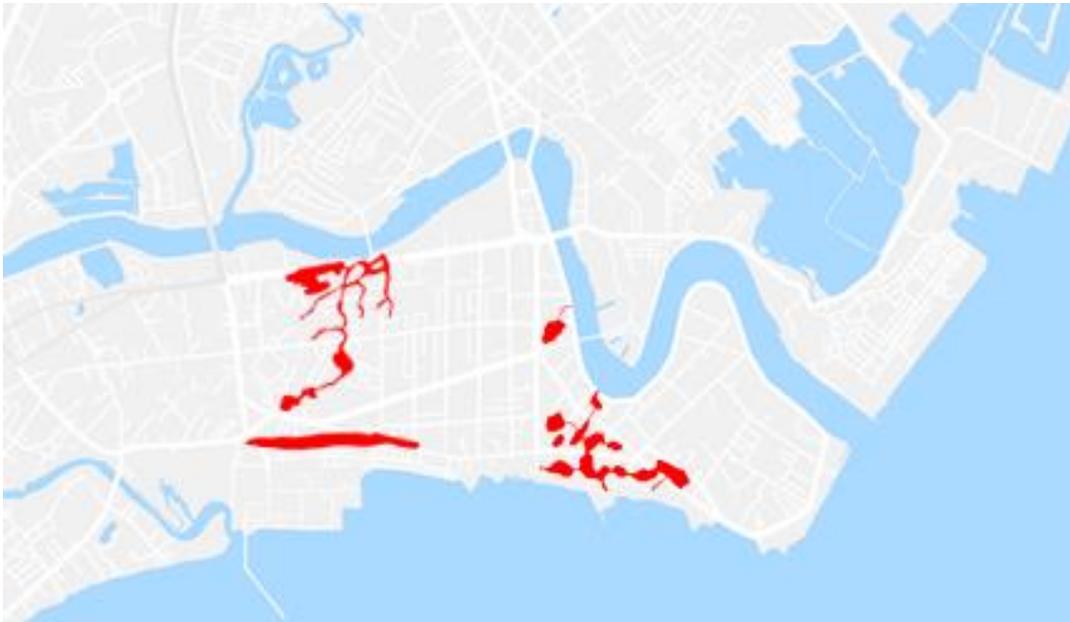

*Figure 8 Vanished bodies of water (red) in Iloilo City.*

### 4.3 Tacloban

Figure 9 shows that there was a pond in the middle of Tacloban City in the year 1944. In the present, this pond is no longer visible. Since the area is an old body of water, locals said that the area is prone to flooding that is why there is also an effort to make proper drainage systems as shown in Figure 10.

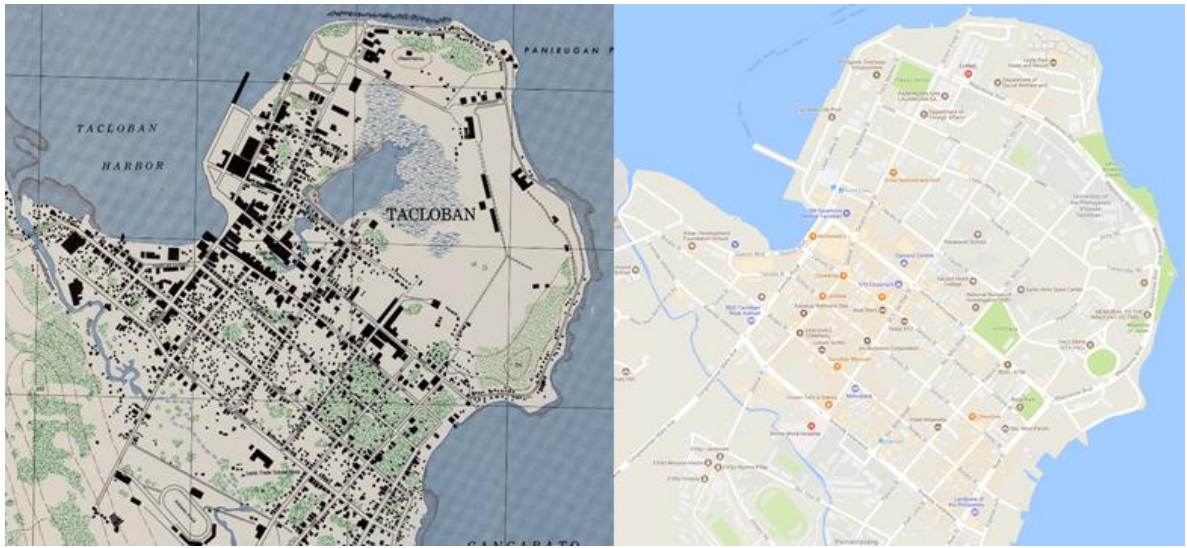

*Figure 9. Left - 1944 map of Tacloban City courtesy of the University of Texas Libraries , Right- corresponding Google Map.*

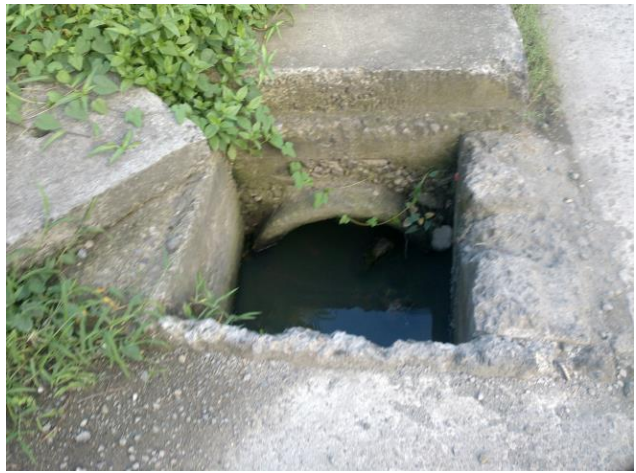

*Figure 10. Part of a drainage system around the former pond area in Tacloban City.*

During the visit, the survey team interviewed locals and stakeholders regarding the suspected reclaimed areas and the missing pond in the city. According to a barangay officer there was an expansion of land area along the shoreline because of the movement of residents towards the sea. The pier (right side of Figure 9 maps ) was also constructed on reclaimed land. A local also confirmed that there was indeed a pond in the center of the city. She claims that she used to swim there with her peers.

While the survey team was scouring local museums and libraries for old Tacloban maps they found a 1940 aerial photograph of Tacloban City (Figure 11) displayed in Hotel Alejandro, one of the oldest hotels in the city. From the photograph, the pond in the middle of the city is clearly visible. Shown in Figure 12 is the trace of the validated vanished pond and reclaimed areas both in the pond and along the shoreline.

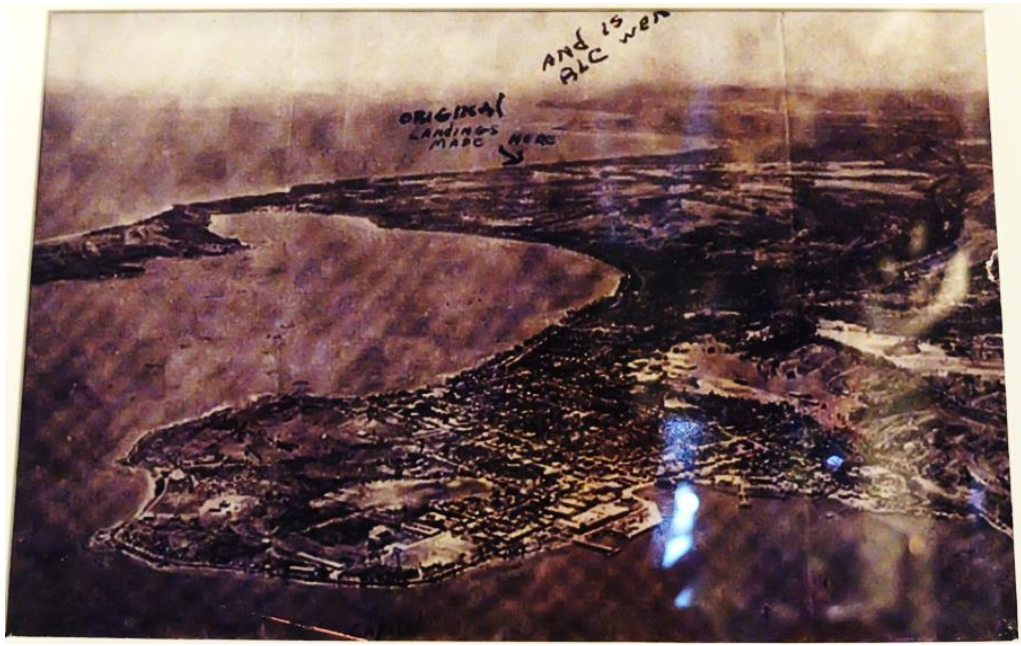

*Figure 11. Vintage aerial photograph of Tacloban City displayed in Hotel Alejandrino. A pond can be clearly seen in the moddle of Tacloban City.*

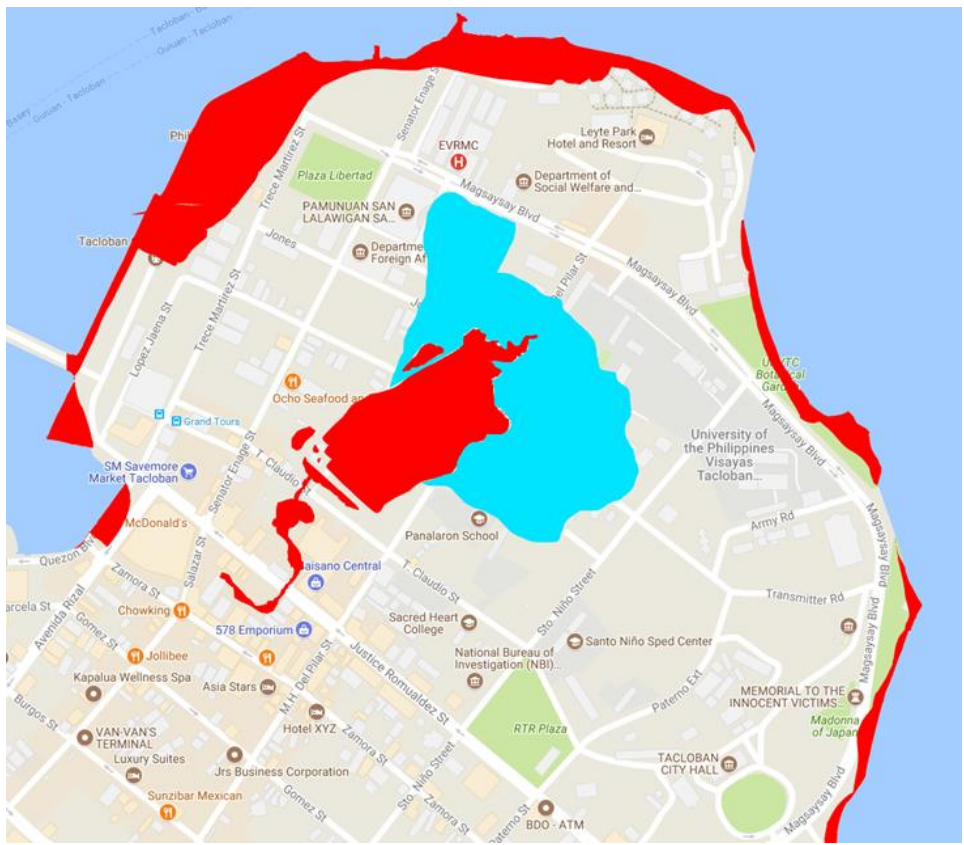

*Figure 12 Trace of validated changes in Tacloban City.  In the interior :  Red - vanished bodies of water Blue - Reclaimed land. Along the coastline : Red – reclaimed land.*

### 4.4 Davao

The missing bodies of water in the Davao maps (shown in Figure 13) are the results of changes in the meandering of the Davao River. Maps from 1944 showed the Davao River flowed towards the eastern part of the city. Presently, the river cut downwards, flowing straight to the sea. The old path of the river dried up and a community was built on top of it. Locals living in that community said that their area is prone to waist-level floods, especially during high tide. Figure 14 shows Google Street View and fieldwork photographs of the same location, in the community built on top of the dried up river. The water remnants indicate the lack of a proper drainage system so flooding has an amplified effect on the community.

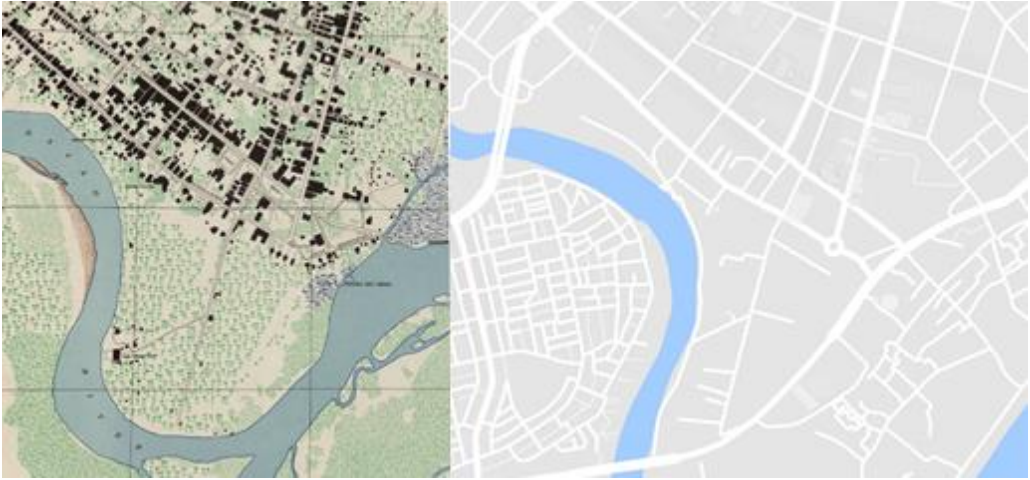

*Figure 13 Left - 1944 map showing Davao River courtesy of the University of Texas Libraries, Right - corresponding Google Map.*

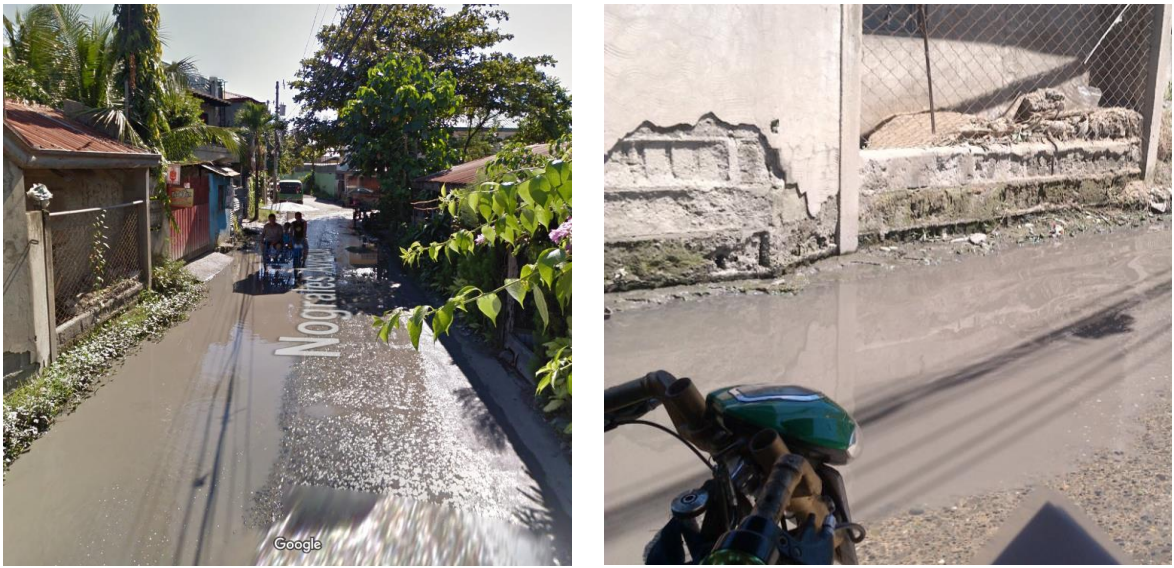

*Figure 14. Left - Google Street View image of a street in Davao City, Right - Photograph of the same location during the validation fieldwork.*

Shown in Figure 15 is the validated path change of Davao River.

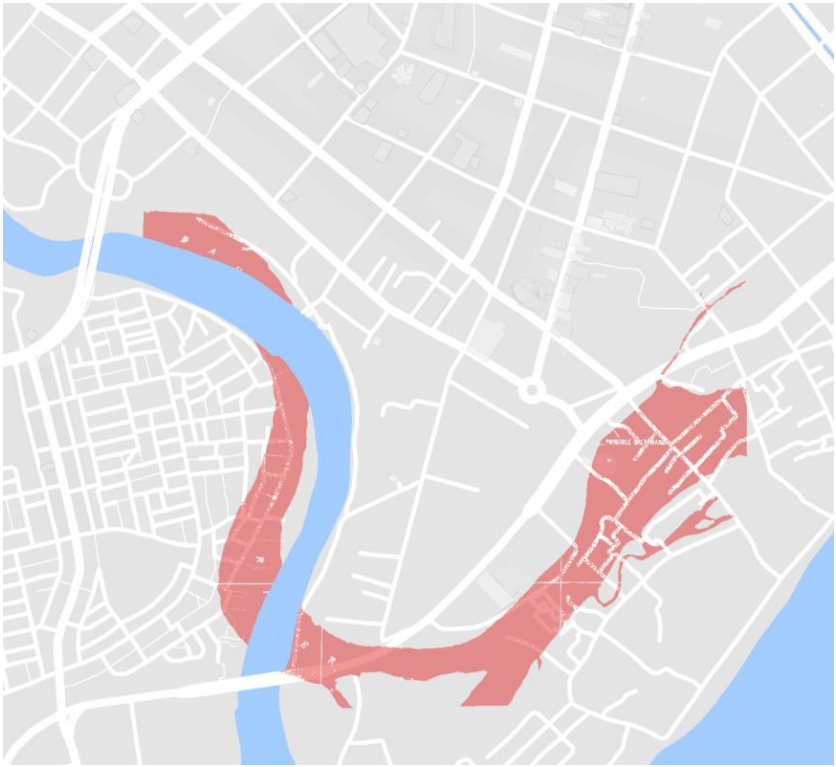

*Figure 15. Red - original path of Davao River. Blue - Current Davao River meander in Google Map*

**4.5 Cebu**

Figure 16 shows three maps of Cebu City. The left map shows a large body of water extending towards the ocean on the right side of the map. In the middle figure , the waterway closed off and formed a lagoon. In Figure 9c, the lagoon disappeared completely. The fieldwork team surveyed the area of the missing lagoon and found out that there is still a body of water (shown in Figure 10) passing through the south side of the original lagoon. A whole community has taken residence in the area of the former lagoon known as Tinago. Interviews with locals stated that their area is lower than sealevel and has a high risk of flooding such that even if it does not rain in their area, if it rains in the mountains, their community gets flooded. A local councilor revealed that her grandmother often tells of the time when they were washing clothes in the lagoon in her younger years.

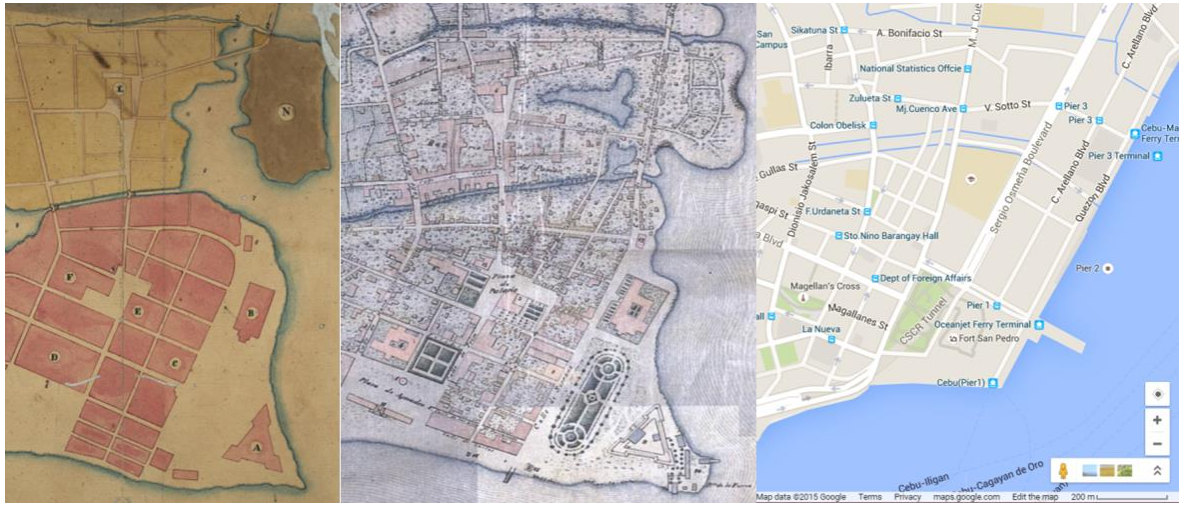

*Figure 16. Georeferenced historical maps of Cebu City. Left - 1833 map courtesy of the National Archives of the Philippines, Middle - 1873 map retrieved online, Right - Google Map.*

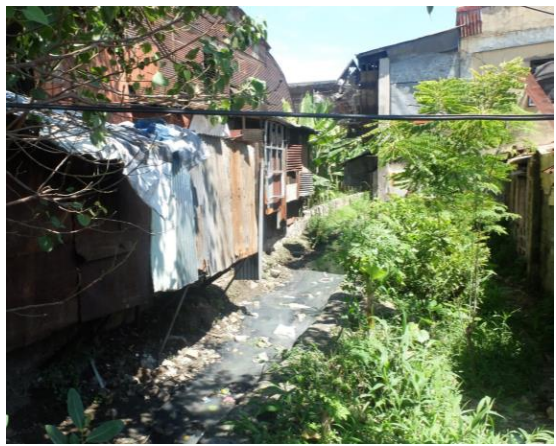

*Figure 17. A small creek on the south side of Tinago. This creek does not appear in Google Maps.*

Google maps.

Figure 18 shows the Parian Estuary in Cebu City. This image shows an example of the reduction of the size of old bodies of water, not necessarily disappearing. The vintage photograph appears to be Parian Estero which used to  wide enough for bancas or small boats to pass through. The image on the right meanwhile is Parian Estero (estuary) at the time of field work. It appears narrower and shallower. Finally, Figure 19 shows the validated traces of lost bodies of water in Cebu City.

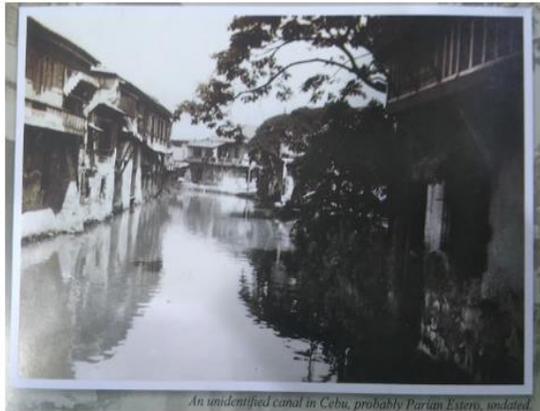
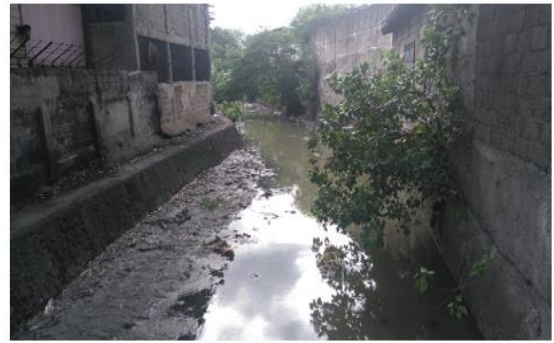

Undated vintage photo from Miller's "Glimpses of Old Cebu"                    March 8, 2017

*Figure 18. An undated photo of an estuary from Miller's "Glimpses of Old Cebu" possibly showing Parian Estuary. Right - Image of the same estuary taken during fieldwork.*

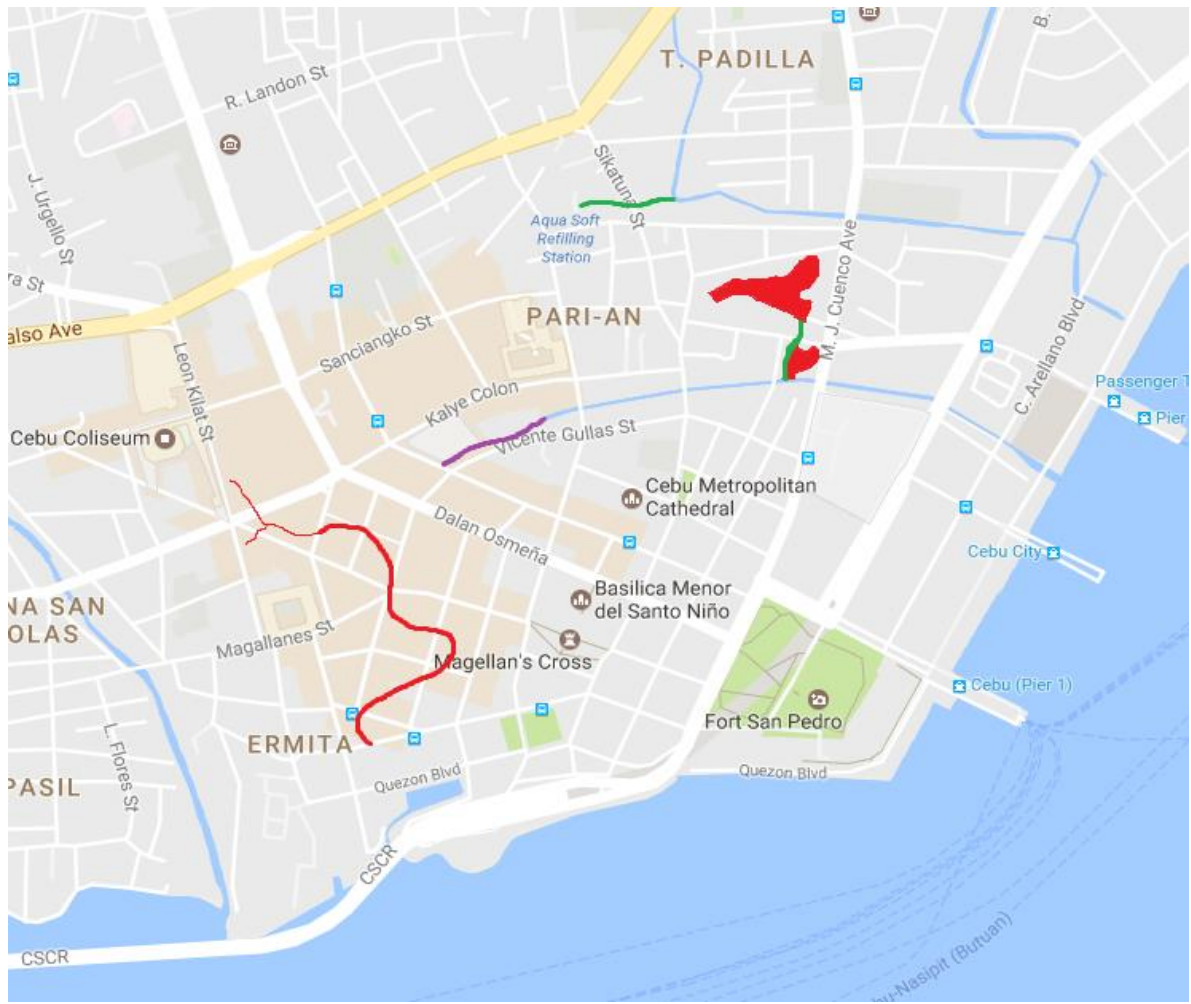

*Figure 19. Red - lost bodies of waters in Cebu City. Green - still existing estuaries. Violet - Estuaries that have gone underground.*

### 4.6 Naga

Figure 20 shows the georectified historical map of Naga City. The fieldwork team confirmed that main parts of the Naga river are still present but the islands within the river that are seen from the historical map are not. The old waterway  jutting from the river going northwards is no longer present. That waterway was believed to be man-made according to Naga City historian Dr. Danilo Gerona and was used to transport goods needed by the bishop.  At the end of this waterway is the Naga Cathedral. Another observation is that the local government already started posting signs (shown in Figure 21) warning the locals about the dangers of the area. The sign directly translates to "Warning! This area is flood prone. Be cautious! Each life is precious."

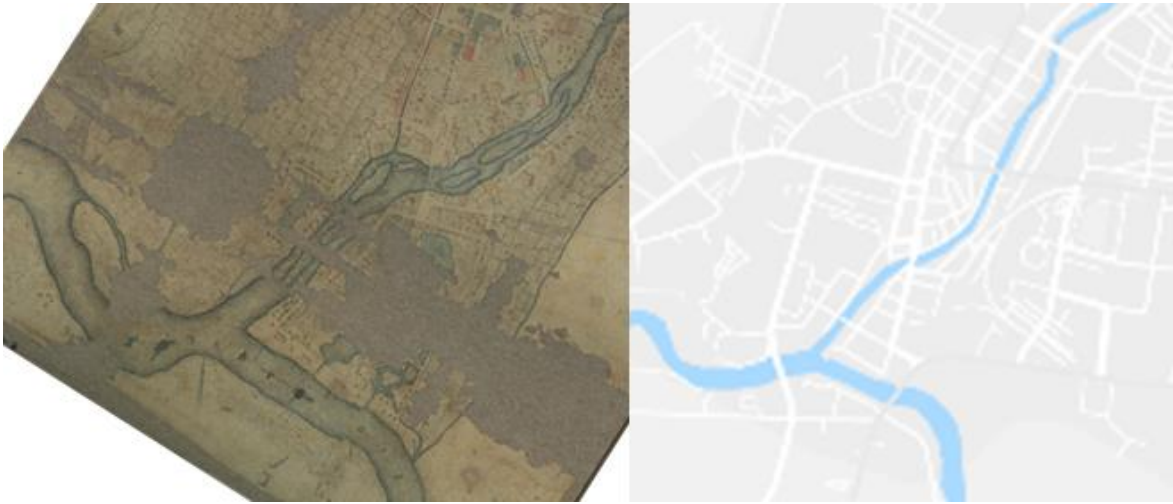

*Figure 20 Left - Georectified 1890 map of Naga City, Bicol Province courtesy of the National Archives of the Philippines. Right - same area in Google Map.*

**Fig. 12.** (a) 1890 historical map and (b) Google map of Naga City.

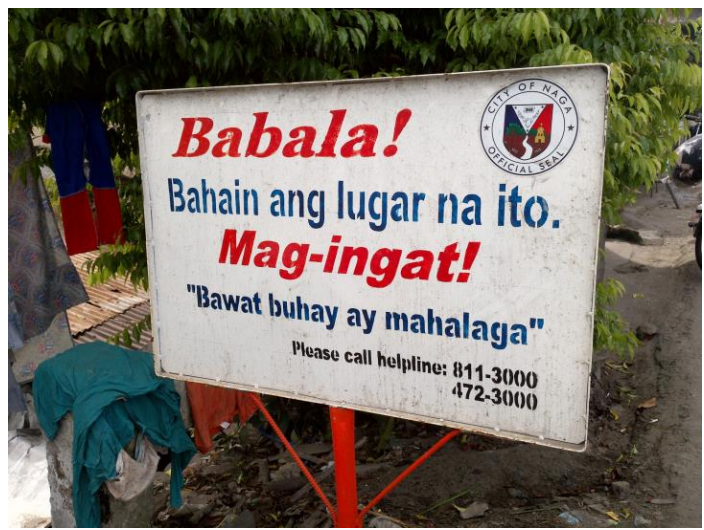

*Figure 21. A sign found in Naga City that directly translates to "Warning! This area is flood prone. Be cautious! Each life is precious.'*

Finally, shown in Figure 22 are the lost and found water ways in Naga City.

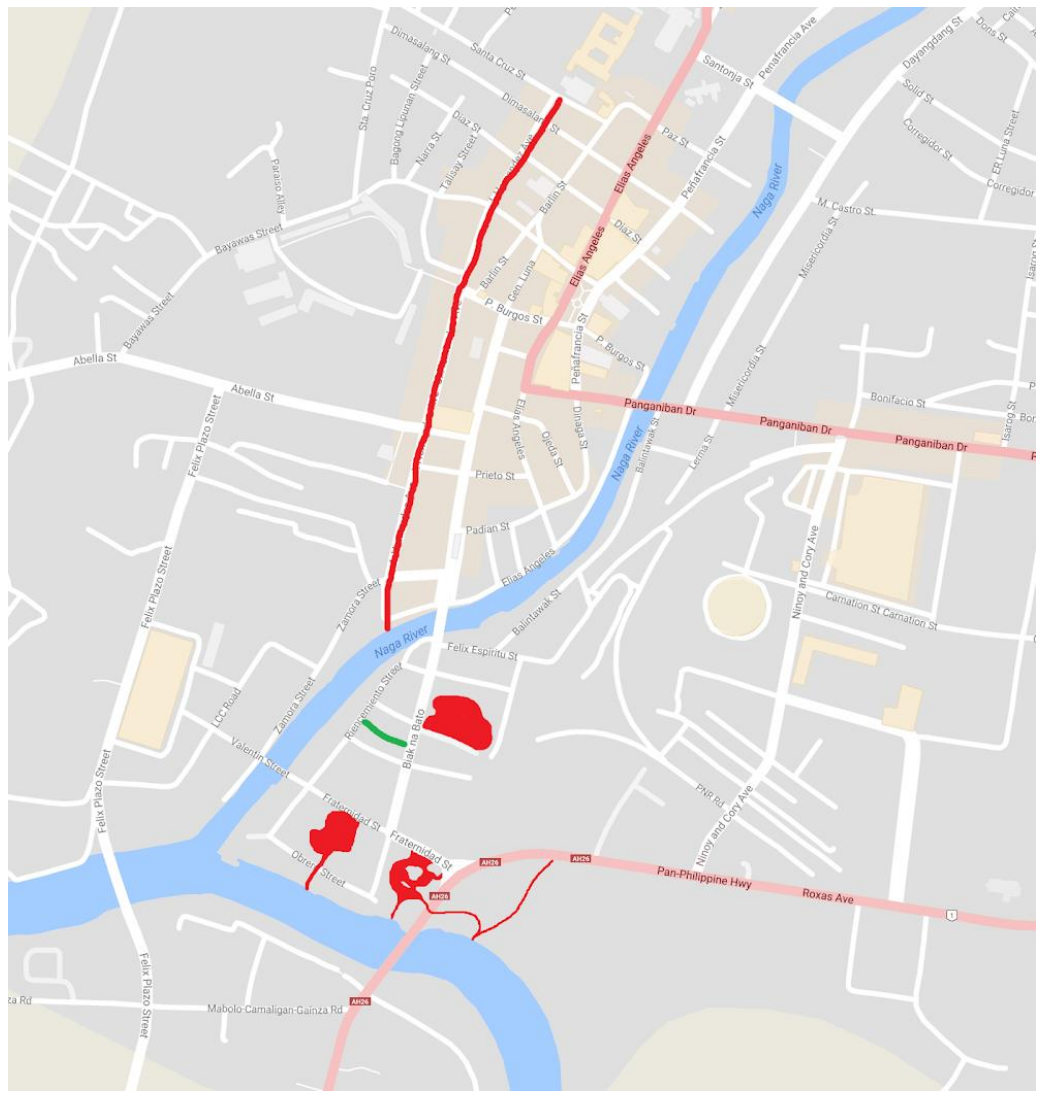

Legend: Disappeared | Still present | Located underground

# 6 Conclusions

By referring to digitized historical maps of Manila City, Iloilo City, Tacloban City, Davao City, Cebu City, and Naga City we were able to identify bodies of water that have shrunk in size or have disappeared altogether. We validated our findings through fieldwork surveys and interviews. We found that depiction of rivers or estuaries in Google Map is not always accurate. Possibly, this is due to the reliance on satellite imagery. Estuaries crowded around by houses or foliage get obscured from top view. As a result, we also report where waterways are actually still present. Interviews with local residents confirm that areas with lost waterways are flood prone.

It was fortunate that in more than one fieldwork the survey team observed ongoing flood control civil works which revealed that there were waterways under built roads. Thus we report the location of waterways that may have been diverted underground.

Vintage photographs of the cities were also sought from local museums and libraries. Interestingly, one should not discount hotels and restaurants because, as what we found in Hotel Alejandro in Tacloban, they can be a source of historical pictures as well.

## Acknowledgements

This work is part of the project "Cartography of Old Informs the New" (COIN) funded by the Department of Science and Technology. Inspiration for this work came from discussions with Prof. Ari Ide-Ektessabi from Kyoto University. The following establishments are also acknowledged for their contribution to the map collection: Filipinas Heritage Library (1875, 1899, and 1908 maps of Manila City), Lopez Museum (1819 map of Manila), National Archives of the Philippines (1883 maps of Iloilo City, 1833 map of Cebu City, 1890 map of Naga City), Army Map Service maps digitized and archived by University of Texas Libraries (1944 maps of Tacloban City and Davao City), and Biblioteca Nacional de España (1893 map of Iloilo).

## References

[01]    Yin, H., & Li, C. (2001). Human impact on floods and flood disasters on the Yangtze River. *Geomorphology*, *41*(2), 105-109.

[02]    Gallanosa, P. B., & Soriano, M. N. (2015, December). Image alignment of historical and current Cebu maps to determine river path changes and flooding pattern. In Humanoid, Nanotechnology, Information Technology, Communication and Control, Environment and Management (HNICEM), 2015 International Conference on (pp. 1-5). IEEE.

[03]    Brown, S. (2013). "The Philippines is the most storm exposed country on Earth," [Online], Available: http://world.time.com/2013/11/11/the-philippines-is-the-most-storm-exposed-country-on-earth/

[04]    Bankoff, G. (2003). Constructing vulnerability: the historical, natural and social generation of flooding in metropolitan Manila. *Disasters*, *27*(3), 224-238.

[05]    Shimizu, Eihan, and Takashi Fuse. "Rubber-sheeting of historical maps in GIS and its application to landscape visualization of old-time cities: focusing on Tokyo of the past." In *Proceedings of the 8th international conference on computers in urban planning and urban management*, vol. 11, no. 1, pp. 3-8. 2003.

[06]    Haase, Dagmar, Ulrich Walz, Marco Neubert, and Matthias Rosenberg. "Changes to Central European landscapes—analysing historical maps to approach current environmental issues, examples from Saxony, Central Germany." Land Use Policy 24, no. 1 (2007): 248-263.

[07]    C. San-Antonio-Gómez, C. Velilla, F. Manzano-Agugliaro, "Urban and landscape changes through historical maps: The Real Sitio of Aranjuez (1775–2005) a case study", Computers Environment and Urban Systems, vol. 44, pp. 47-58, 2014.

[08]    G. Timár, B. Székely, G. Molnár, C. Ferencz, A. Kern, C. Galambos, C. Galambos, G. Gercsák, L. Zentai, "Combination of historical maps and satellite images of the Banat region-Re-appearance of an old wetland area", Global and Planetary Change, vol. 62, no. 1, pp. 29-38, 2008.

[09]    S.D. Laycock, P.G. Brown, R.G. Laycock, A.M. Day, "Aligning archive maps and extracting footprints for analysis of historic urban environments", Computers & Graphics, vol. 35, no. 2, pp. 242-249, 2011.